\documentclass[sigconf,screen]{acmart}
 
\usepackage{booktabs} 

\copyrightyear{2018} 
\acmYear{2018} 
\setcopyright{acmlicensed}
\acmConference[SWAN '18]{Proceedings of the 4th ACM SIGSOFT International Workshop on Software Analytics}{November 5, 2018}{Lake Buena Vista, FL, USA}
\acmBooktitle{Proceedings of the 4th ACM SIGSOFT International Workshop on Software Analytics (SWAN '18), November 5, 2018, Lake Buena Vista, FL, USA}
\acmPrice{15.00}
\acmDOI{10.1145/3278142.3278143}
\acmISBN{978-1-4503-6056-2/18/11}

\graphicspath{{figures/}}
\DeclareGraphicsExtensions{.pdf,.png}

\usepackage{xcolor}

\usepackage[export]{adjustbox} 
\usepackage{amssymb,amsmath}

\usepackage{tabularx}

\usepackage{enumitem}

\usepackage{calc}

\usepackage{multirow}

\usepackage{amsthm}
\newtheoremstyle{definitiontight}
  {.5\baselineskip\@plus.2\baselineskip\@minus.2\baselineskip}
  {.6\baselineskip\@plus.2\baselineskip\@minus.2\baselineskip}
  {}
  {}
  {\bfseries}
  {.}
  {.5em}
  {}
\theoremstyle{definitiontight}
\newtheorem{definition}{Definition}[section]
\newtheorem{hypothesis}{Hypothesis}[section]

\usepackage{amsmath}
\usepackage{amssymb} 
\usepackage{eucal} 

\usepackage{siunitx}
\definecolor{VeryLightGray}{rgb}{0.92,0.92,0.92}
\usepackage{colortbl}

\usepackage{subfig}

\usepackage{balance}

\begin{document}
\title[(No) Influence of Continuous Integration on the Commit Activity in...]{(No) Influence of Continuous Integration on the Commit~Activity~in~GitHub~Projects}

\author{Sebastian Baltes}
\orcid{0000-0002-2442-7522}
\affiliation{%
  \institution{University of Trier}
  \city{Trier}
  \country{Germany}
}
\email{research@sbaltes.com}

\author{Jascha Knack, Daniel Anastasiou, Ralf Tymann}
\affiliation{%
  \institution{University of Trier}
  \city{Trier}
  \country{Germany}
}

\author{Stephan Diehl}
\orcid{0000-0002-4287-7447}
\affiliation{%
  \institution{University of Trier}
  \city{Trier}
  \country{Germany}
}
\email{diehl@uni-trier.de}

\renewcommand{\shortauthors}{S. Baltes, J. Knack, D. Anastasiou, R. Tymann, and S. Diehl}

\begin{abstract}
A core goal of Continuous Integration (CI) is to make small incremental changes to software projects, which are integrated frequently into a mainline repository or branch.
This paper presents an empirical study that investigates if developers adjust their commit activity towards the above-mentioned goal after projects start using CI.
We analyzed the commit and merge activity in 93 GitHub projects that introduced the hosted CI system \emph{Travis CI}, but have previously been developed for at least one year before introducing CI.
In our analysis, we only found one non-negligible effect, an increased merge ratio, meaning that there were more merging commits in relation to all commits after the projects started using \emph{Travis CI}.
This effect has also been reported in related work.
However, we observed the same effect in a random sample of 60 GitHub projects not using CI.
Thus, it is unlikely that the effect is caused by the introduction of CI alone.
We conclude that: (1) in our sample of projects, the introduction of CI did not lead to major changes in developers' commit activity, and (2) it is important to compare the commit activity to a baseline before attributing an effect to a treatment that may not be the cause for the observed effect.
\end{abstract}

%
%



\maketitle

\section{Introduction}
\label{sec:introduction}

Continuous Integration (CI) is a software engineering practice where developers frequently integrate their work into a common ``mainline'' repository or branch~\cite{Fowler2006, Meyer2014}.
After changes have been committed to this repository or branch, a CI system automatically builds and tests the software. 
The approach has originally been proposed by Grady Booch~\cite{Booch1991} and became popular after being promoted as one of the Extreme Programming (XP) practices~\cite{Beck2000}.
Many software projects on GitHub use hosted CI services such as \emph{Travis CI} or \emph{CloudBees}~\cite{GousiosZaidmanOthers2015}.
The specific CI process may differ between projects (e.g., when is a build triggered, what is considered to be a broken build, etc.)~\cite{StahlBosch2014}, but a core goal of CI is to work with small increments and to integrate them frequently into the mainline branch~\cite{Meyer2014}.
There has been research on different aspects of using CI in GitHub projects (see Section~\ref{sec:related-work}) and also one study investigating if the introduction of CI actually leads to a different commit activity in terms of smaller, but more frequent commits~\cite{ZhaoSerebrenikOthers2017}.
That study found an increasing number of merge commits after the introduction of CI, but a large variation regarding the ``commit small'' guideline.
We assessed the CI guidelines with a different methodological lens and also found an increased merge ratio, that is more merging commits in relation to all commits after the projects started using CI.
However, we observed the same effect in a random sample of projects not using CI and conclude that it is unlikely that the effect is caused by the introduction of CI alone.

\begin{figure}[b]
\centering
\includegraphics[width=0.9\columnwidth,  trim=0.0in 0.2in 0.0in 0.0in]{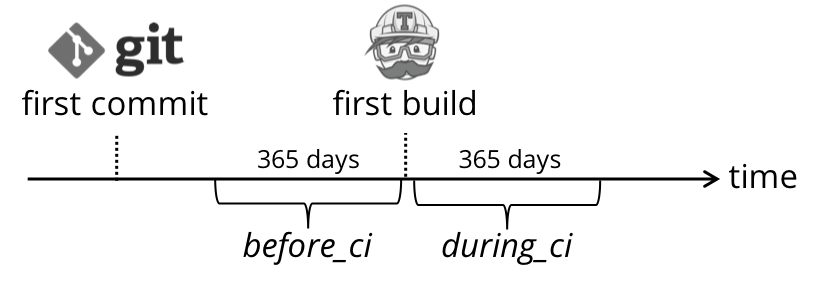} 
\caption{Partitioning of commits into two time frames for each analyzed project. We only analyzed the activity one year before and after the first build.}
\label{fig:timeline}
\end{figure}

\section{Research Design}
\label{sec:research-design}

The overall goal of our research was to analyze the impact of the introduction of continuous integration on the commit and merge activity in open source software projects.
We selected projects hosted on GitHub that employ the CI system \emph{Travis CI}.
We identified such projects using the \emph{TravisTorrent data set} (January 11, 2017)~\cite{BellerGousiosOthers2017} and the \emph{GHTorrent} Big Query data set (February 14, 2017)~\cite{Gousios2013, Gousios2018}.
We only considered projects that:

\begin{enumerate}
\item were active for at least one year (365 days) before the first build with \emph{Travis CI} (see \emph{before\_ci} in Figure~\ref{fig:timeline}),
\item used \emph{Travis CI} at least for one year (see \emph{during\_ci} in Figure~\ref{fig:timeline}),
\item had commit or merge activity on the default branch in both of these phases, and
\item used the default branch to trigger builds.
\end{enumerate}

The motivation for restrictions (1) and (2) was to be able to compare a considerable amount of commit activity before and after the first build.
To further exclude projects that use a different branch than the default branch as ``mainline'', we added restrictions (3) and (4).

We utilized the \emph{GHTorrent} Big Query data set to identify the time frames \emph{before\_ci} and \emph{during\_ci} for the projects in the \emph{TravisTorrent} data set.
Of all projects in the data set, 544 satisfied restrictions (1) and (2).
Of these projects, 366 were Ruby projects and 178 were Java projects.
The mean time span \textit{before\_ci} was 2.9 years ($SD{=}2.1$, $Mdn{=}2.2$),
the mean time span \textit{during\_ci} was 3.2 years ($SD{=}1.1$, $Mdn{=}3.2$).
To compare two time frames of equal size, we restricted our analysis to the activity one year before and after the first build.

Our \emph{units of observation} were the commits in the selected projects.
We only considered changes to Java or Ruby source code files, because we were interested in the actual development activity, not in changes to binary files like images or PDF documents.
We cloned all 544 project repos and extracted the version history for all branches with a tool we developed~\cite{Baltes2017}.
For each repo and branch, we created one log file with all regular commits and one log file with all merges.
From those log files, we then extracted metadata about the commits and stored this data in CSV files using a second tool~\cite{Baltes2017b}.
Combining the commit data and \emph{TravisTorrent}, we applied restrictions (3), and (4).
Moreover, we excluded projects where the first commit activity happened more than seven days before the project creation date according to \emph{GHTorrent}.
The motivation for this additional filtering step was that projects which started outside of GitHub and have later been ported could introduce a bias, because the commit activity may differ if features such as GitHub pull requests are not available~\cite{GousiosStoreyOthers2016, TsayDabbishOthers2014, GousiosPinzgerOthers2014}.
This resulted in a sample of 113 projects (89 Ruby and 24 Java projects).

As \emph{units of analysis}, we considered the commits themselves and the projects. 
The measures we compared for these units were the \emph{commit rate}, the \emph{commit size}, and the \emph{merge ratio}, which we define below.

In the following, let $\mathcal{C}$ be the set of all commits in all projects we collected as described above.
Further, let $\mathcal{P}_\mathcal{C}$ be the power set of $\mathcal{C}$.
The \textit{commit frequency} is the number of commits in a certain time span.
We chose a time span of one week to adjust for the varying activity between working days and weekend.
The partition $C^{w} = C^{w_1} \dots C^{w_n}$ divides a set of commits $C$ into one subset for each week, beginning with the week of the first commit ($w_1$) until the week of the last commit ($w_n$) in $C$.
For our analysis, we ignored $w_1$ or $w_n$ if they did not contain data about a whole week.

Considering only non-merging commits, 52\% of those weeks were inactive, meaning there was no commit activity.
For the merging commits, there were 71\% inactive weeks.
A possible reason for this phenomenon could be developers working intensively on an open source project for a few days and then focusing on something else, for example other closed- or open-source projects.
We consider an investigation of activity patterns in open source software projects to be an interesting direction for future work.
In our analysis, we focused on active weeks weeks and excluded weeks without any commit activity.
 
In a last filtering step, we excluded projects without merges in either of the two phases or only data for incomplete weeks.
This resulted in a final set of 93 projects.
We provide the project list and the scripts and data used for the filtering process as supplementary material~\cite{BaltesKnack2018, Baltes2018}.
With the terminology introduced above, we can now define the \emph{median commit rate} based on a weekly partition of a set of commits $C$:

\begin{definition}[Commit Rate]
Let $C \in \mathcal{P}_\mathcal{C}$ be a set of commits and $C^{w} = C^{w_1} \dots C^{w_n}$ be a weekly partition of $C$.
We define the \textit{median commit rate} $\overline{c_{w}} \colon \mathcal{P}_\mathcal{C} \to \mathbb{R}_{0}^{+}$ as:
\[ \overline{c_{w}}(C) = \textit{median}(|C^{w_i}|), \text{ } i \in \{1,  \dots, n \}  \]
\end{definition}

Please not that, since we focus on active weeks in our analysis, a project with ten commits in one week and no commits in the nine following weeks has a higher commit rate than a project with one commit per week spread over 10 weeks.
Beside the commit rate, we define two measures to describe the \textit{change size of a commit}.
In the following, $n_f(c)$ denotes the number of source code files changed by a commit $c \in C$, and $n_l(c)$ denotes the number of lines changed in the source code files modified by this commit.
We define the code churn $n_l(c)$ as $l_c^+ + l_c^-$, where $l_c^+$ is the sum of all lines added to the modified files and $l_c^-$ is sum of all lines deleted from those files. Both $l_c^+$ and $l_c^-$ are based on the line-based diffs of the modified files.
Please note that this measure overestimates the change size in case existing lines are modified, e.g., a change to one existing line is represented as one deleted and one added line.

\begin{definition}[Change Size]
For a commit $c \in \mathcal{C}$, we define the \textit{change breadth} $b \colon \mathcal{C} \to \mathbb{R}_{0}^{+}$ and the \textit{change depth} $d \colon \mathcal{C} \to \mathbb{R}_{0}^{+}$ as:
\[ b(c) =  n_f(c)
\hspace{1em} and \hspace{1em} 
d(c) = \frac{n_l(c)}{n_{f}(c)} \]
For a set $C \in \mathcal{P}_\mathcal{C} $ of commits, we define the \textit{median change breadth} $~\overline{b}(C)~$ as:
\[ \overline{b}(C) = \textit{median}(b(c)), \text{ } c \in C \]
For a partition $C^\phi = \{C^1 \dots C^n$\} of $C$, we define the \textit{median change breadth} $~\overline{b}(C^\phi)~$ as:
\[ \overline{b}(C^\phi) = \textit{median}(\overline{b}(C^i)), \text{ } i \in \{1, \dots, n\} \]
The \textit{median change depths} $~\overline{d}(C)~$ and $~\overline{d}(C^\phi)~$ are defined analogously.
\end{definition}

The last measure we are going to define is the amount of merging commits in relation to all commits.
We use this measure to estimate the branching and merging activity in a certain time span. 
For all $C \in \mathcal{P}_\mathcal{C}$, let $n_m(C)$ denote the number of commits in $C$ that merged other commits.

\begin{definition}[Merge Ratio]
For a set $C \in \mathcal{P}_\mathcal{C} $ of commits, we define the \textit{merge ratio} $m \colon \mathcal{P}_\mathcal{C} \to [0, 1]$ as:
\[ m(C) = \frac{n_m(C)}{|C|}\]
\end{definition}

Now that we have precise definitions of \textit{commit rate}, \textit{change size}, and \textit{merge ratio}, we can take different comparison perspectives to describe the commit activity in software projects.
We have two parameters for aggregating single commits into sets of commits that we can then compare with our measures.
First, we can aggregate commits according to their \emph{origin}, e.g., all commits in the same project or from the same developer.
Second, we can aggregate commits according to \emph{time}, here mainly the two time frames before and after the first CI build.
Regarding the \emph{origin} of the commits, we define two partitions:

\begin{itemize}[labelindent=\parindent, labelwidth=\widthof{$C^\textit{p,d}:$}, label=$C^\textit{p,d}:$, leftmargin=*, align=parleft, parsep=0pt, partopsep=0pt, topsep=1ex, noitemsep]
\item[$C^\textit{p}:$] Partitions $C$ into one set of commits for each project $p \in \textit{projects}(C)$.
\item[$C^\textit{p,b}:$] Partitions $C$ into one set of commits for each branch in each project ($\forall p \in \textit{projects}(C) \text{ } \forall b \in \textit{branches}(p)$).
\end{itemize}

For this paper, we do not consider other partitions such as $C^{p,d}$ for each developer in a project.
Regarding the \emph{time}, we define two partitions that correspond to the time frames described above:

\begin{itemize}[labelindent=\parindent, labelwidth=\widthof{$C^\textit{p}_\textit{during}$}, label=$C^\textit{p}_\textit{during}$, leftmargin=*, align=parleft, parsep=0pt, partopsep=0pt, topsep=1ex, noitemsep]
\item[$C_\textit{before}$] Contains all commits before the first CI build for each project, but not more than 365 days before the first build.
\item[$C_\textit{during}$] Contains all commits after the first but before the last CI build for each project, but not more than 365 days after the first build. 
\end{itemize}

It is possible to combine the above-mentioned partitions.~$C_\textit{before}^\textit{p, b}$, for instance, contains one set of commits for each branch $b$ in each project $p$, taking only commits into account that were committed before the first build in the corresponding project.
For the following analyses, we first partition the sets using the branch partition $C^\textit{p, b}$ and only consider the default (``mainline'') branch for each project.
This branch is often, but not always, called \texttt{master}.
For better readability, we only write $C^\textit{p}$ instead of $~C^{p, b},~b=\textit{default\_branch}(p)$ in the following.
We focused on the default branch, because this is usually the branch triggering the CI builds.
In fact, according to \emph{TravisTorrent}, 81.8\% of all builds of the 93 projects in our final sample were triggered by commits to the default branch.
For the following analyses, we take a \emph{project perspective}, i.e., we partition the \textit{before} and \textit{during} sets into the commits for each project, then calculate and compare $\overline{c_w}$, $\overline{b}$, $\overline{d}$, and $m$ for each project.



\section{Results}
\label{sec:results}


In the following, we test our a priori hypotheses about commit activity changes using the data we collected.  
We provide the raw data and all analysis scripts as supplementary material~\cite{BaltesKnack2018, Baltes2018}.

As descriptive statistics, we report median ($Mdn$), interquartile range ($IQR$), and mean ($M$).
To test for significant differences, we applied the non-parametric two-sided \textit{Wilcoxon signed rank test}~\cite{Wilcoxon1945} and report the corresponding p-values ($p_w$, significance level $\alpha=0.01$).
To measure the effect size, we used \textit{Cliff's delta} ($\delta$)~\cite{Cliff1993}.
Its values range between $+1$, when all values of one group are higher than the values of the other group, and $-1$, when the reverse is true.
We also provide the confidence interval of $\delta$ at a 95\% confidence level  ($CI_\delta$).
Our interpretation of $\delta$ is based on the thresholds described by Romano et al.~\cite{RomanoKromreyOthers2006}: \emph{negligible effect} ($|\delta|<0.147$), \emph{small effect} ($0.147\le|\delta|<0.33$), \emph{medium effect} ($0.33\le|\delta|<0.474$), otherwise \emph{large effect}.
Our general assumption was that, with the introduction of CI, the projects shift towards smaller increments (in form of smaller commits) that are frequently integrated (directly committed or merged) into the default branch. 

\subsection{Commit Rate}

If developers follow the advice to frequently integrate their work into the main branch, the introduction of CI may have an effect on the commit rate: 

\begin{hypothesis}
After the introduction of CI, the commit rate increases.
\end{hypothesis}

\noindent To test this hypothesis, we compared the following sets:

\begin{align*}
\{ \overline{c_\textit{w}}(C^\textit{p}_\textit{before}) \} &\;\textit{vs.}\; \{ \overline{c_\textit{w}}(C^\textit{p}_\textit{during}) \},~p \in \textit{projects}(C)\\
\end{align*}

For $C_\textit{before}$, the median commit rate per project was $2.5$ commits per week ($IQR=2.5$, $M=3.5$); for $C_\textit{during}$ it was $2.0$ commits per week ($IQR=2$, $M=3.3$).
The difference was not significant ($p_w = 0.66$) and negligible ($\delta=-0.02$, $CI_\delta=[-0.18, 0.15]$).

The median commit rate of merging commits was $1$ merge per week ($IQR=1$, $M=1.5$) in the timespan \textit{before} and also $1$ merge per week ($IQR=1$, $M=1.5$) in the timespan \textit{during}.
The difference was was not significant ($p_w = 0.10$) and negligible  ($\delta=0.09$, $CI_\delta=[-0.06, 0.23]$).

Thus, we failed to reject the null hypothesis that the commit rate decreases or does not change.

\begin{figure*}
\centering
\subfloat[][Merge ratio (number of merging commits in relation to all commits).]{
	\includegraphics[width=0.48\textwidth,  trim=0.1in 0.6in 0.1in 0.1in]{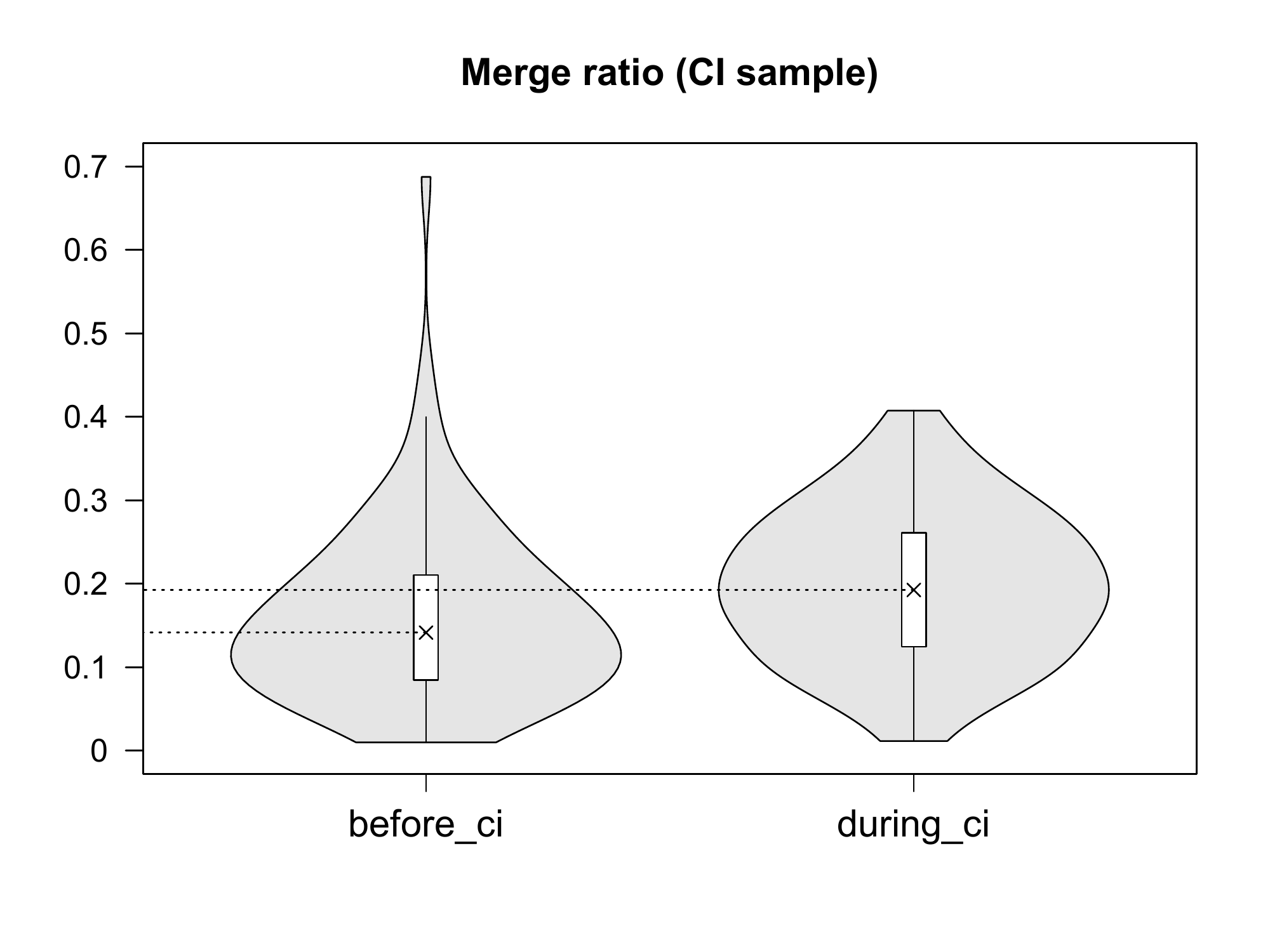} 
	\label{fig:merge-ratio}
}
\hfill
\subfloat[][Pull request ratio (number of commits merging a pull request in relation to all merging commits).]{
	\includegraphics[width=0.48\textwidth,  trim=0.1in 0.6in 0.1in 0.1in]{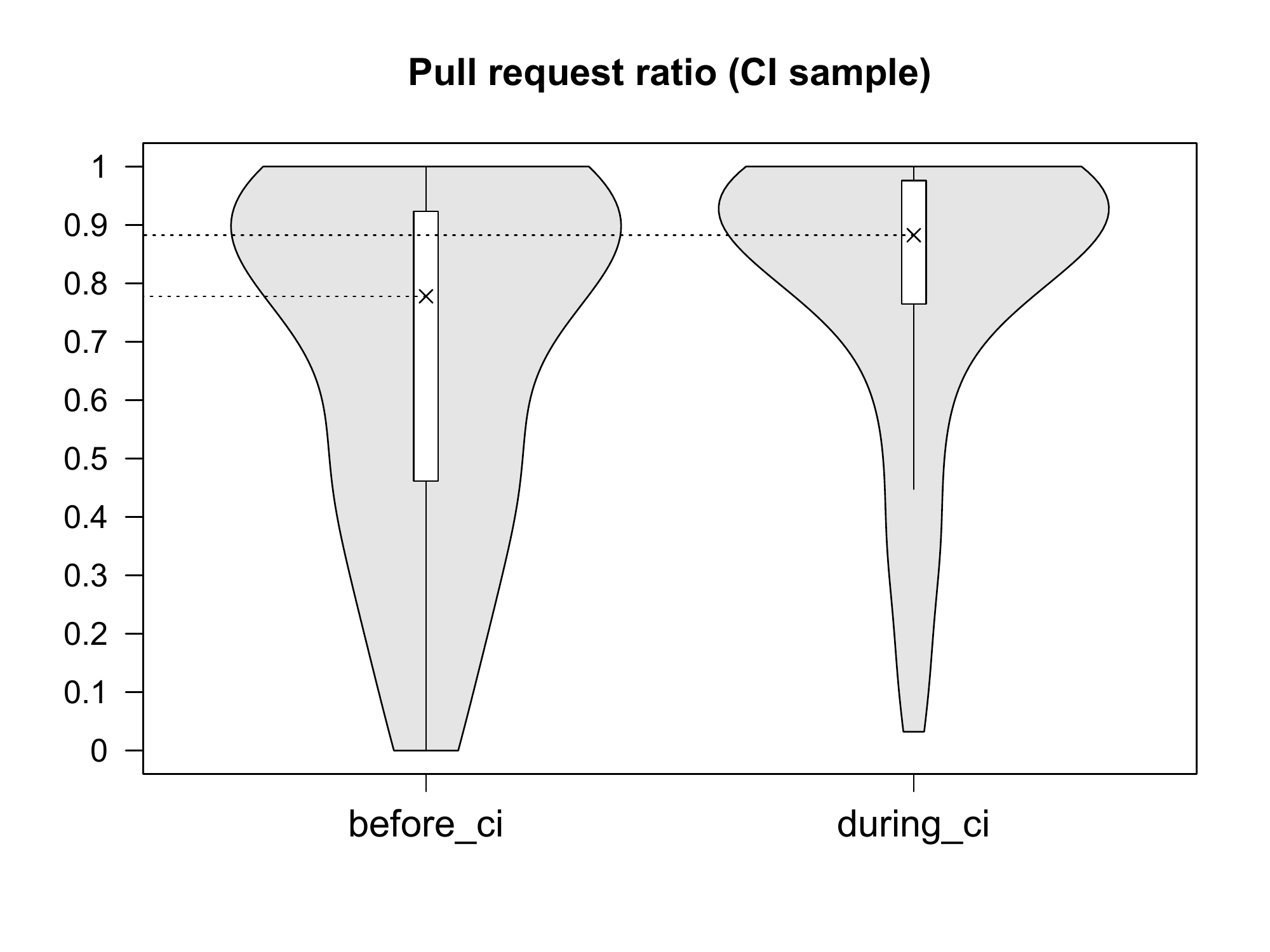}  
	\label{fig:pr-ratio}
}
\caption{Merge ratio and pull request ratio of projects ($n=93$) before and after their first CI build.}
\end{figure*}

\subsection{Change Size}

As described above, we employed two measures to capture the change size of a commit.
We tested both its breadth, i.e., the number of changed source code files, as well as its depth, i.e., the number of changed lines divided by the number of files.
Our hypothesis was that, with the introduction of CI, the median size of the changes decreases:

\begin{hypothesis}
After the introduction of CI, the commit changes become smaller, i.e., they have a lower change breadth and depth.
\end{hypothesis}

\noindent To test this hypothesis, we compared the following sets:

\begin{align*}
\{ \overline{b}(C^\textit{p}_\textit{before}) \} &\;\textit{vs.}\; \{ \overline{b}(C^\textit{p}_\textit{during}) \},~p \in \textit{projects}(C) \\
\{ \overline{d}(C^\textit{p}_\textit{before}) \} &\;\textit{vs.}\; \{ \overline{d}(C^\textit{p}_\textit{during}) \},~p \in \textit{projects}(C) \\
\end{align*}

For $C_\textit{before}$, the median commit breadth per project was $1$ file per commit ($IQR=1$, $M=1.8$); for $C_\textit{during}$ it was also $1$ file per commit ($IQR=1$, $M=1.6$).
The difference was not significant ($p_w = 0.05$) and negligible ($\delta=-0.07$, $CI_\delta=[-0.21, 0.06]$).
The median commit breadth of merged commits was $2$ files per merged commit ($IQR=1$, $M=2.8$) in the timespan \textit{before} and also $2$ files per merged commit ($IQR=0$, $M=2.0$) in the timespan \textit{during}.
The difference was was not significant ($p_w = 0.04$) and negligible  ($\delta=-0.10$, $CI_\delta=[-0.24, 0.06]$).

For $C_\textit{before}$, the median commit depth per project was $7.8$ lines per file ($IQR=4.8$, $M=9.9$); for $C_\textit{during}$ it was $7.0$ lines per file ($IQR=4.3$, $M=8.6$).
The difference was significant ($p_w = 0.0008$), but negligible ($\delta=-0.10$, $CI_\delta=[-0.26, 0.06]$).
The median commit depth of merged commits was $10.0$ lines per file ($IQR=9.4$, $M=16.8$) in the timespan \textit{before} and $9.5$ lines per file ($IQR=7.5$, $M=12.4$) in the timespan \textit{during}.
The difference was was not significant ($p_w = 0.10$) and negligible ($\delta=-0.08$, $CI_\delta=[-0.24, 0.09]$).

Thus, we failed to reject the null hypothesis that the change size decreases or does not change.

\subsection{Merge Ratio}

As a last step, we analyzed if the introduction of CI increases the merge ratio, and in particular the pull request ratio (the number of commits merging a pull request in relation to all merging commits).
Our hypothesis was that after the introduction of CI, the amount of direct commits in the default branch decreases, because developers prefer to review the changes before triggering a build.
On GitHub, the pull-based software development model has become more and more popular over the past years~\cite{VasilescuYuOthers2015, GousiosPinzgerOthers2014, GousiosStoreyOthers2016, GousiosZaidmanOthers2015}.
This development model allows projects to review the changes before merging pull requests into the default branch.

\begin{hypothesis}
After the introduction of CI, the merge ratio increases.
\end{hypothesis}

\noindent To test this hypothesis, we compared the following sets:

\begin{align*}
\{ m(C^\textit{p}_\textit{before}) \} &\;\textit{vs.}\; \{ m(C^\textit{p}_\textit{during}) \},~p \in \textit{projects}(C) \\
\end{align*}

For $C_\textit{before}$, the median merge ratio over all projects was $0.14$ ($IQR=0.13$, $M=0.16$); for $C_\textit{during}$ it was $0.19$ ($IQR=0.14$, $M=0.20$).
The difference was significant ($p_w<\num{5.4e-06}$) and the effect was small ($\delta=0.28$, $CI_\delta=[0.12, 0.43]$).
Thus, we reject the null hypothesis that the change size decreases or does not change.
Figure~\ref{fig:merge-ratio} shows violin plots visualizing the difference.

We also compared the pull request ratio:
For $C_\textit{before}$, the median pull request ratio was $0.78$ ($IQR=0.46$, $M=0.68$); for $C_\textit{during}$ it was $0.88$ ($IQR=0.21$, $M=0.80$).
The difference was significant ($p_w<\num{7.1e-05}$) and the effect was small ($\delta=0.24$, $CI_\delta=[0.08, 0.40]$).
Thus, an increased usage of pull requests is likely to be one major factor leading to the increased merge ratio.

\subsection{Comparison Sample}

To check whether the increased merge ratio can actually be attributed to the introduction of CI, we analyzed how the merge ratio changed in a random sample of GitHub projects that do not use CI.
Since this sample should be comparable to the CI project sample, we applied the following constraints, selecting projects that:


\begin{enumerate}
\item have Java or Ruby as their project language
\item have commit activity for at least two years (730 days)
\item are engineered software projects (at least 10 watchers)
\end{enumerate}

We applied those constraints to the projects in the \emph{GHTorrent BigQuery data set} (February 06, 2018)~\cite{Gousios2013, Gousios2018}.
Moreover, we used the same filter as for the CI projects to remove projects with commits more than one week before the GitHub project creation date.
Since all projects in the CI sample were ``engineered software projects''~\cite{MunaiahKrohOthers2017}, we applied filter (3) to exclude small ``toy'' GitHub projects~\cite{KalliamvakouGousiosOthers2014}.
We applied this popularity filter, using the number of watchers or stargazers, because it has been used in several well-received studies and proved to have a very high precision in selecting engineered projects~\cite{MunaiahKrohOthers2017} (almost 0\% false positives for a threshold of 10 watchers/stargazers).
All of the 93 projects in the CI sample satisfied this constraint. 

Of the 8,405 projects that satisfied the constraints, 359 were also in the \emph{TravisTorrent} data set.
We excluded those projects and drew a random sample of 800 projects from the remaining 8,046 projects.
We retrieved the commit data in the same way as for the CI projects.
Then, we determined the date that splits the development activity in those projects into two equally-sized time frames.
For the analysis, we considered one year (365 days) of commit activity in the default branch before and after that date.
We only considered projects with commit and merge activity in the default branch in both time frames (130 projects), which we manually checked for CI configuration files.
We removed 70 projects with such configuration files, resulting in 60 projects in which we did not find any indication that they use CI services.

For those projects, we compared the merge ratio in the two time frames to investigate if we can observe a similar increase like in the CI projects.
For the commits before the split date, the median merge ratio over all projects was $0.12$ ($IQR=0.17$, $M=0.16$); for the commits after that date it was $0.19$ ($IQR=0.28$, $M=0.22$).
The difference between the merge ratios of all projects in the two time frames was significant ($p_w = 0.006$) and the effect was small ($\delta=0.20$, $CI_\delta=[-0.008, 0.40]$).
As this effect was similar to the CI sample, we cannot attribute the increased merge ratio exclusively to the introduction of CI.

\begin{figure}
\centering
\includegraphics[width=1\columnwidth,  trim=0.1in 0.6in 0.1in 0.1in]{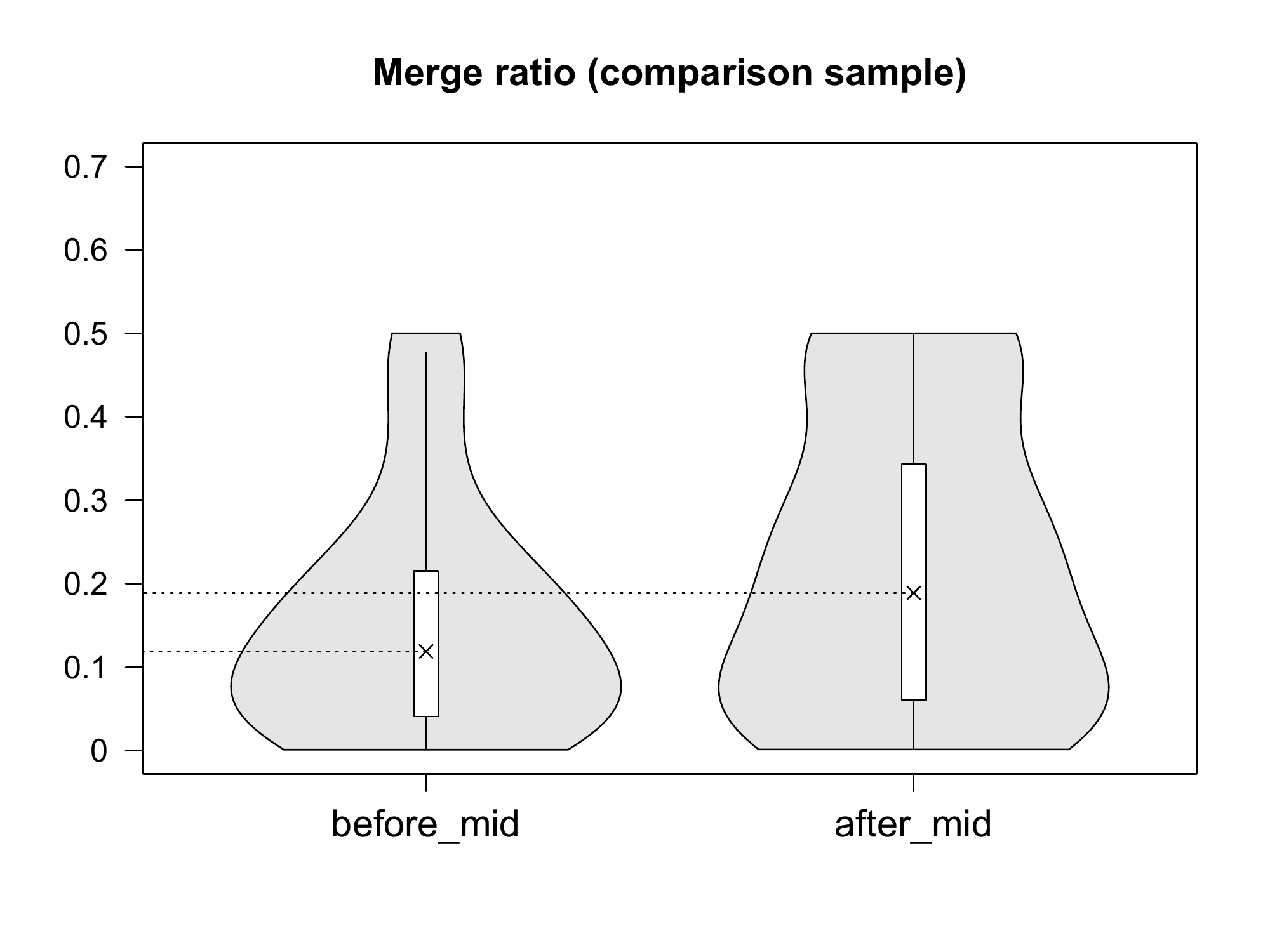} 
\caption{Merge ratio in randomly selected projects not using CI ($n=60$) before and after median development date.}
\label{fig:merge-ratio-random}
\end{figure}

\section{Limitations and Verifiability}
\label{sec:limitations+verifiability}

The main limitations of our research are the focus on Ruby and Java projects and the fact that we do not know if the introduction of CI would actually be the trigger for the effects we hypothesized.
In fact, we are confident that the increased merge ratio cannot be attributed to the introduction of CI alone.
Another limitation is the fact that projects may have used other CI tools than \emph{Travis CI} before, meaning that our time span \textit{before\_ci} would actually be \textit{during\_ci}, but with another CI tool.
This could be one reason why we did not observe the expected effects.

To check whether the higher merge ratio after the introduction of CI depends on the number of project contributors, we compared projects with many contributors to projects with fewer contributors.
First, we identified all project contributors using the committer and author email addresses in the commit metadata we collected, yielding a median number of 14 contributors per project ($IQR=13$, $M=21.8$).
We split the CI project sample into two groups, one with 47 small ($\leq 14$ contributors) and one with 46 large ($> 14$ contributors) projects.
Then, we compared the merge ratios before and after the introduction of CI for those two groups separately.
In both groups, the merge ratio was significantly higher after the introduction of CI ($p_w < 0.005$) and the effect was small (small projects: $\delta=0.29$, $CI_\delta=[0.06, 0.48]$; large projects: $\delta=0.28$, $CI_\delta=[0.06, 0.49]$).
Thus, we conclude that the increased merge ratio is independent from the number of project contributors.

In our random comparison sample of projects not using CI, we compared the commit activity before and after the median date.
Obviously, the effects may be different when choosing another comparison date.
Moreover, we used a weekly perspective.
Other time frames such as months (see Zhao et al.'s work in Section~\ref{sec:related-work}) could lead to different results.
Because of the common distinction into five workdays and two days weekend, we think a weekly perspective is reasonable.

We excluded projects with a commit history outside of GitHub (see Section~\ref{sec:research-design}).
However, projects may have been imported with one large initial commit, without importing their complete Git history.
Even if such a huge commit would be part of the timespan \textit{before}, it is unlikely to bias our results, because our comparison is based on median and interquartile range and we applied robust statistical methods~\cite{KitchenhamMadeyskiOthers2017}.

Our focus on open source GitHub projects limits the transferability of our results to closed source commercial projects.
In those projects, the number of weeks without commit activity is likely to be much lower than in open source projects, which are often not developed full-time.
We consider a comparison of the effects of introducing CI in open source versus closed source projects to be an important direction for future work.

To enable other researcher to verify our results, we provide the raw data and our analysis scripts as supplementary material~\cite{BaltesKnack2018, Baltes2018}.

\section{Related Work}
\label{sec:related-work}

Vasilescu et al.~\cite{VasilescuYuOthers2015} analyzed GitHub projects which introduced CI and found that CI improves the productivity of project teams in terms of merged pull requests.
Hilton et al.~\cite{HiltonTunnellOthers2016}, who analyzed GitHub projects, \emph{Travis CI} builds, and surveyed 442 developers, found that (1) popular projects are more likely to use CI, (2) projects that use CI release more than twice as often as those that do not use CI, and (3) CI builds on the master branch pass more often than on the other branches.
Furthermore, Hilton et al.~\cite{HiltonNelsonOthers2017} conducted semi-structured interviews with developers and conclude that developers encounter increased complexity, increased time costs, and new security concerns when working with CI.

Zhao et al.~\cite{ZhaoSerebrenikOthers2017} investigated the impact of CI on GitHub projects.
Their approach was similar to ours as they also considered one year of activity before and after the first CI build.
However, they aggregated information for a whole month, opposed to one week in our analysis, and utilized a statistical modeling framework named \emph{regression discontinuity design}~\cite{ImbensLemieux2008}.
Like us, they observed an increased number of merge commits over time, but as we described above, this trend is not limited to projects adapting CI.
Thus, we argue against attributing this effect to the introduction of CI.

Zhao et al. also conclude that the adaption of CI is much more complex than suggested by other studies.
For example, they found that more pull requests are being closed after adopting CI, but their analysis suggests that the expected increasing trend over time manifests itself only before adopting CI, afterwards the number of closed pull requests remains relatively stable.
This indicates that more work is needed to investigate if and how projects adapt after interventions such as the introduction of CI.

Memon et al.~\cite{MemonGaoOthers2017} analyzed data from Google's CI system \emph{TAP} and found that code recently modified by more than three developers is more likely to break the build.
Other studies investigated aspects such as the usage of static code analysis tools in CI pipelines~\cite{ZampettiScalabrinoOthers2017}, the personnel overhead of CI~\cite{ManglavitiCoronadoMontoyaOthers2017}, the interplay between non-functional requirements and CI builds~\cite{PaixaoCriciaOthers2017}, the impact of CI on developer attraction and retention~\cite{GuptaKhanOthers2017} or code reviews~\cite{RahmanRoy2017}, and factors influencing build failures~\cite{RauschHummerOthers2017, BellerGousiosOthers2017b, IslamZibran2017, ReboucasSantosOthers2017, ZiftciReardon2017}.

\section{Conclusion and Future Work}
\label{sec:conclusion}

We presented an empirical study investigating if developers of open-source GitHub projects adjust their commit activity towards smaller but more frequent commits after the introduction of continuous integration (CI).
We expected this change, because a core goal of CI is to work with small incremental changes.
We analyzed the commit and merge activity in 93 GitHub projects that introduced \emph{Travis CI} and have been developed on GitHub for at least one year before the introduction of CI.
The only non-negligible effect we observed was an increased merge ratio, i.e., the number of merging commits in relation to all commits.
However, we observed the same effect in a random sample of GitHub projects that do not use CI.
Thus, we argue against attributing this effect to the introduction of CI alone.
It is more likely that projects use merges more frequently when they grow and mature.
Another reason could be the general trend of adopting the pull-based software development model~\cite{VasilescuYuOthers2015, GousiosPinzgerOthers2014, GousiosStoreyOthers2016, GousiosZaidmanOthers2015}.

Beside those findings, this paper contributes a precise formalization of different commit activity measures, which we used to test for expectable changes in GitHub projects after the introduction of CI.
Our results show that it is important to compare observed changes in commit activity to a baseline (in our case the comparison sample) to prevent attributing those changes to a treatment that may not be the actual cause.

Directions for future work include analyzing the commit activity from additional comparison perspectives, for example by partitioning the commits per developer as mentioned in Section~\ref{sec:research-design}.
Moreover, one could broaden the research by conducting a more holistic quantitative analysis to identify dominant factors causing the increased merge ratio, for example using multiple regression.
Another way to broaden the research would be to continue with a qualitative analysis, asking GitHub developers how they perceive the impact of CI on their projects.

\begin{acks}
The authors would like to thank Oliver Moseler for his feedback on our formalization of the commit activity measures.
Moreover, we thank the anonymous reviewers for their helpful comments.
\end{acks}


\balance
\bibliographystyle{ACM-Reference-Format}
\bibliography{literature}

\end{document}